\documentstyle[prl,aps,epsf,multicol]{revtex}
\begin{document}
\draft

\def\i{\imath\,}
\def\ih{\frac{\imath}{2}\,}
\def\undertext#1{\vtop{\hbox{#1}\kern 1pt \hrule}}
\def\ra{\rightarrow}
\def\lfa{\leftarrow}
\def\ua{\uparrow}
\def\da{\downarrow}
\def\Ra{\Rightarrow}
\def\lra{\longrightarrow}
\def\ler{\leftrightarrow}
\def\lrb#1{\left(#1\right)}
\def\O#1{O\left(#1\right)}
\def\EV#1{\left\langle#1\right\rangle}
\def\tr{\hbox{tr}\,}
\def\trb#1{\tr\lrb{#1}}
\def\dd#1{\frac{d}{d#1}}
\def\dbyd#1#2{\frac{d#1}{d#2}}
\def\pp#1{\frac{\partial}{\partial#1}}
\def\pbyp#1#2{\frac{\partial#1}{\partial#2}} 
\def\pd#1{\partial_{#1}}
\def\br{\\ \nonumber & &}
\def\brr{\right. \\ \nonumber & &\left.}
\def\inv#1{\frac{1}{#1}}
\def\be{\begin{equation}}
\def\ee{\end{equation}}
\def\bea{\begin{eqnarray}}
\def\eea{\end{eqnarray}}
\def\ct#1{\cite{#1}}
\def\rf#1{(\ref{#1})}
\def\EXP#1{\exp\left(#1\right)} 
\def\INT#1#2{\int_{#1}^{#2}} 
\def\LHS{left-hand side }
\def\RHS{right-hand side }
\def\COM#1#2{\left\lbrack #1\,,\,#2\right\rbrack}
\def\AC#1#2{\left\lbrace #1\,,\,#2\right\rbrace}

\title{Fractionalization in the cuprates: Detecting the topological order}
\author{T. Senthil and Matthew P.A. Fisher}
\address{
Institute for Theoretical Physics, University of California,
Santa Barbara, CA 93106--4030
}

\date{\today}
\maketitle

\begin{abstract}
The precise theoretical characterization of a fractionalized phase
in spatial dimensions higher than one is through the concept of 
``topological order''. We describe a physical effect that is a 
robust and direct consequence of this hidden order that should enable a 
precise experimental characterization of fractionalized phases. In particular,
we propose specific ``smoking-gun'' experiments to unambiguously settle the 
issue of electron fractionalization in the underdoped cuprates.
  
\end{abstract}
\vspace{0.15cm}


\begin{multicols}{2}
\narrowtext
 
Does the electron splinter apart ({\em i.e} fractionalize) in the cuprate
high-$T_c$ materials? This question has been the subject of heated
debate for some years, and is apparently far from being settled. 
Phenomenologically,
electron fractionalizion\cite{PWA,KRS,LN,z2short} is
very appealing, primarily
because it provides a simple explanation of
the superconductivity.  Upon fractionalization the charge of the
electron is liberated from it's Fermi statistics, 
thereby allowing the electron charge to {\it directly} condense
leading to superconductivity without
invoking ideas of pairing. Further, several other unusual properties of 
the cuprates find 
a natural explanation in terms of the fractionalization idea, most notably 
the angle-resolved photoemission results\cite{z2short}.

Despite these attractive features, there has been considerable difficulty
in constructing an acceptable theory of fractionalization for the cuprates.
This is partly
due to inadequacies in 
our theoretical understanding,
but more problematic has been 
the lack of clear experimental signatures indicative
of electron fractionalization.
In this paper, building on our recent theoretical
understanding of fractionalization using a $Z_2$
gauge theory formulation\cite{z2long},
we overcome this difficulty. In particular, we 
predict a novel physical effect that is 
a robust property of fractionalized phases -
this makes possible a direct and unambiguous experimental test of whether 
the electron splinters apart in the cuprates.

{\bf Fractionalization theory}:
In our recent work\cite{z2long} we demonstrated that a general class of
strongly interacting electron models could be recast in the form  
of a $Z_2$ gauge theory, which then enabled us to provide a 
reliable discussion of issues of electron fractionalization.
In particular, we demonstrated the possibility of obtaining fractionalized 
phases in two or higher spatial dimensions. In such a phase, the electron splits
into two independent excitations - the spin of the electron is carried by a neutral
fermionic excitation (the ``spinon'') and the 
charge is carried by a bosonic excitation
(the ``chargon'')\cite{note1}.   
There is a third distinct excitation, 
namely the flux of the 
$Z_2$ gauge field (dubbed the ``vison''). The vison is gapped in the fractionalized phase. 
The $Z_2$ gauge theory approach is closely related 
to the ideas on vortex pairing\cite{NLII} 
and other gauge theoretic formulations\cite{SNL} of fractionalization.

The precise theoretical characterization of a fractionalized  
phase is in it's ``topological order''\cite{Wen1,z2long}
- a concept that has been elucidated clearly in the context of the
quantum Hall effect by Wen and coworkers\cite{Wen2}. 
A fractionalized phase in a manifold with a 
non-trivial topology has a ground state degeneracy
which depends on the topology.  A vison that is trapped in each ``hole'' 
in the manifold stays there forever,
but does not affect the ground state energy in the thermodynamic limit. 
Consider for instance a cylinder. There are two states 
depending on whether or not a vison has threaded the cylinder. 
The inability of the trapped vison to escape from the cylinder 
is the hallmark of the 
fractionalized phase.
 
The topological order inherent in a fractionalized phase
endows it with a tremendous amount of robustness to various
``real-life'' complications.
For example, topological order survives\cite{z2long} in the presence of weak 
amounts of disorder. It can also coexist with various other conventional 
broken symmetries, such as charge or spin ordering\cite{NLII,z2long,z2short}. 
Thus, if the cuprates do show
fractionalization, features like charge stripes or even antiferromagnetism are
side effects (albeit interesting ones) and not directly related to 
the origin of the superconductivity.

It is also important to understand the role of a 
finite non-zero temperature  
on the topological order. Theoretically, the effect of finite temperature
depends crucially\cite{z2long} on the spatial dimension. In two spatial
dimensions (2d), the visons are point-like excitations, and are gapped
in the fractionalized phase.  But at any non-zero temperature, there will
be a finite density of thermally excited visons, thereby
destroying the topological order.
In a three dimensional fractionalized phase, however, 
the visons are loop-like excitations, costing 
a finite energy per unit length.  Consequently, at low temperatures 
large vison loops are absent and the topological order survives.  
As the temperature is increased there will eventually be a true phase
transition where the vison loops unbind. 

For quasi-two dimensional layered materials
such as the cuprates,
two distinct fractionalized phases are possible. First, the
fractionalization can occur independently in each layer with the different
layers being ``decoupled" from one another. In this case, 
vison loops proliferate {\it between} the layers, and the topological 
order exists only for visons which penetrate a layer.  As in 2d,
this order
is destroyed by
arbitrarily small temperature.  Alternatively, 
with strong enough interlayer coupling it is possible
that the interlayer vison loops are also expelled,
resulting in a
phase with full three dimensional topological order that
survives at
low non-zero temperatures.

{\bf Detecting the topological order}: Armed with the theoretical 
understanding described above, we 
may pose a sharp and definite question about the cuprates.   
Quite generally, there are three qualitatively distinct possibilities for the 
behavior of the underdoped cuprates 
with regard to the phenomenon of fractionalization. 

\begin{figure}
\epsfxsize=3.3in
\centerline{\epsffile{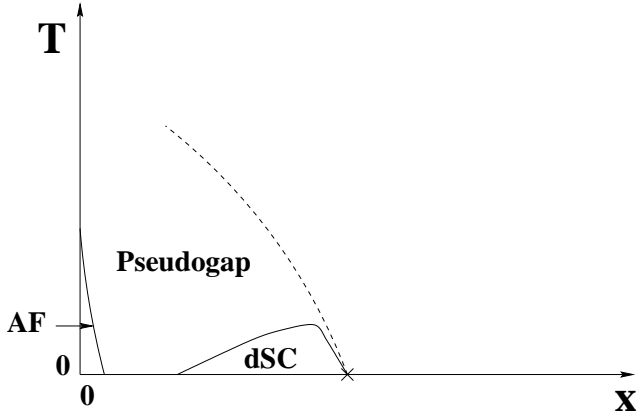}}
\vspace{0.15in}
\caption{One of the three possibilities for the underdoped cuprates 
with regard to electron fractionalization and the associated topological order:
There simply is no topological order}
\vspace{0.05in}
\label{noto}
\end{figure}

(i) Fractionalization and the associated 
topological order simply does not occur in the cuprates (see Fig. \ref{noto}). 

\begin{figure}
\epsfxsize=3.3in
\centerline{\epsffile{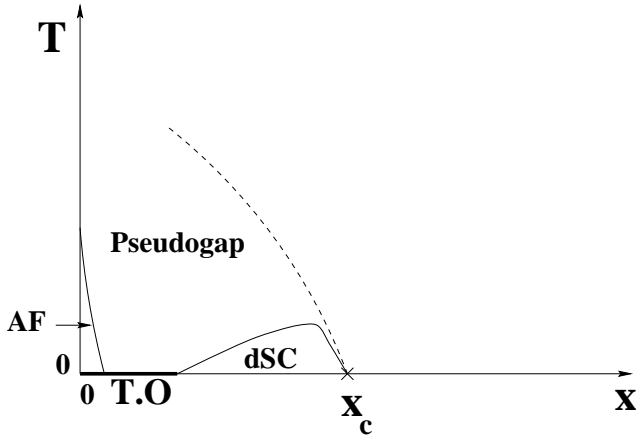}}
\vspace{0.15in}
\caption{Quasi-$2d$ fractionalization: The topological order(T.O) is
strictly present only at zero temperature. The dashed line 
describing the pseudogap crossover corresponds to the crossover to the 
$T = 0$ fractionalized phase. }
\vspace{0.05in}
\label{to2d}
\end{figure}

(ii)The topological order occurs independently in each 
two dimensional layer, and hence strictly speaking, exists only at zero
temperature (see Fig. \ref{to2d}).
   
\begin{figure}
\epsfxsize=3.3in
\centerline{\epsffile{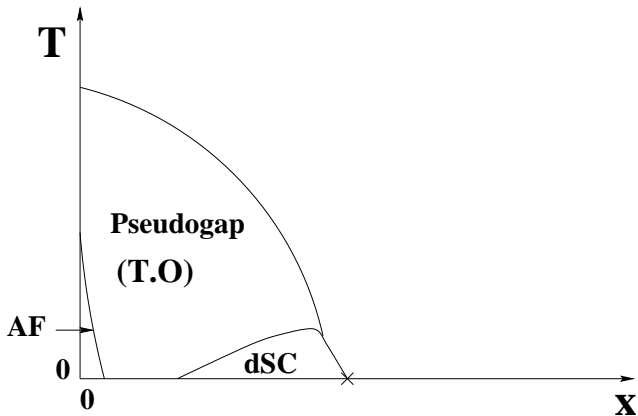}}
\vspace{0.15in}
\caption{$3d$ fractionalization: The topological order survives upto a finite 
non-zero temperature.}
\vspace{0.05in}
\label{to3d}
\end{figure}

(iii)The topological order is three dimensional, and 
survives up to a non-zero temperature (see Fig. \ref{to3d}).

Which one of these three possibilities is actually realized 
in the real materials? 
It is widely believed that the pseudogap line in the cuprates is only a 
crossover and not a true phase transition. This, and other 
phenomenological considerations (specifically
the ``incoherent" c-axis transport), probably disfavor 3d 
topological order as a serious possibility in the cuprates. 
Nevertheless, for conceptual purposes it will be very useful 
to begin by considering this case.    
Though this is distinguished from the other 
two cases by having a finite temperature
phase transition, a {\em direct} experimental 
characterization of the 3d topological order would be preferable. 

We now propose an experiment
that is directly sensitive 
to the presence of topological order, presuming initially
that scenario (iii) is realized.   
Imagine the following sequence of events:

(a) Start with an underdoped sample
in a cylindrical geometry, with the axis of the cylinder
perpendicular to the layers.
In the presence of a magnetic field,
cool into the superconducting phase
such that exactly 
one $hc/2e$ flux quantum is trapped in the hole of the cylinder. 

(b) Heat the sample to above $T_c$. {\em The magnetic flux penetrates into the sample, 
but the vison will still
be trapped}. This is because of the finite temperature
topological order in case (iii). 

(c) Now turn off the magnetic field. The vison will still remain trapped. 
We have thereby
prepared the sample above $T_c$ with zero magnetic field
in a state with a vison threading the 
cylinder. 

(d) How do we tell that there is a vison trapped? The simple way is 
to cool the sample back down below $T_c$. 
The trapped vison still cannot escape 
but {\it must} nucleate an $hc/2e$ quantum of magnetic flux. 
This flux will be generated {\it spontaneously}
and can be in either direction -  
thereby breaking the time reversal invariance achieved in (c)!

An alternate experiment is to again repeat the sequence of events (a) to (d), 
but now work at a fixed very low temperature
and move from the superconductor into 
the (underdoped) insulator, and back, by adiabatically 
tuning some parameter. Again, one
should see a spontaneous $hc/2e$ flux generated if 
the ground state on the insulating side 
is fractionalized. 
This experiment is, of course, much more challenging. 
It has recently been demonstrated however
that an electrostatic field\cite{Ahn} can be used to 
move across the superconductor-insulator phase boundary at low temperature,
at least for very thin films. 
Another possibility 
is to use photodoping (or perhaps even pressure).

These experiments offer a direct and conceptually straightforward way
to detect the presence of 3d
topological order. 
In particular, if case (i) is what is actually realized, there 
will certainly be no spontaneous flux
generated at the end of either experiment. 

But what if
the topological order is two-dimensional
(case (ii)), as seems more likely for the cuprates?
In the experiment performed by tuning 
across the superconductor-insulator phase boundary
at {\it zero} temperature, a
spontaneous flux is certainly expected.
But at
finite temperature the topological
order is strictly speaking absent, 
and a trapped vison
will  
ultimately escape at long times. 
As is usual in such cases, the result will then depend on how fast 
the experiment is performed.
In step (b) above, 
the magnetic flux penetrates into the sample almost immediately
but at low temperatures the vison will remain trapped for a much longer time
of order 
$t_v \sim t_0 e^{\frac{E_0}{k_B T}}$.  Here, $E_0$ is the energy
cost for a vison in a single layer and 
$t_0$ is
some microscopic time scale. Provided the time scale of the 
experiment is smaller than $t_v$, a 
spontaneous magnetic flux should be present
at the end.  But if the time scale of 
the experiment is significantly longer
than $t_v$, there will be no flux and the 
presence of the ($T=0$) topological order will be missed.
It is therefore 
essential that the experiment be performed over time scales
smaller than the vison decay time. 
To this end, it is clearly advantageous to heat the sample
only slightly above $T_c$ in step (b) above.
Interlayer coupling will presumably
enhance $t_v$, so that 
less anisotropic materials might also be preferable.

An advantage of these experiments is that they offer
a {\it direct} and unambiguous route to detect the
presence of topological order.  With topological order,
the spontaneous generation of flux should be insensitive to 
unavoidable materials complications, such as impurities
and other coexisting broken symmetries (stripes, etc). 
In that sense, this experiment
gives a characterization of
fractionalization that is as clear-cut as the Meissner effect is for 
superconductivity, or quantization of the Hall conductivity for the quantum 
Hall effect.

A number of other equally robust predictions can be made when the
conditions in the above experiment are slightly modified.
Specifically, if the initial
magnetic flux trapped in step (a) is $hc/e$ rather than $hc/2e$, 
there will be no trapped vison in
step (b) - hence there will be no spontaneous flux observed. 
More generally, if
the initial flux is an odd multiple of $hc/2e$, a spontaneous flux of 
$hc/2e$ will be generated at the end,
whereas with an 
even multiple of $hc/2e$ initially, there will be no spontaneous flux.
This even/odd effect is a direct reflection of the 
Ising character of electron fractionalization above one dimension.

Another interesting geometry to consider, particularly if case (ii)
is realized, is to repeat the experiment with the axis of the 
cylinder parallel to the layers. Then, the vison can escape
by passing between adjacent layers,
and no spontaneous flux
will be generated (independent of the initial flux).

Once a vison is trapped in a topologically ordered phase, one can 
also imagine other more subtle physical effects
that will distinguish it from the same sample with no 
trapped vison. For instance, thermal or 
even electrical conductance of a cylinder with a 
trapped vison will be different from 
that without the vison, since the interference contribution 
from paths of spinons or chargons that wind around 
the cylinder will be affected (extra minus sign for 
an odd winding if vison is present).  Detecting this effect
will of course require ``phase-coherence'' around the cylinder - 
which presumably only occurs
at very low temperatures and in a small
sample. In the cuprates with $d$-wave pairing, thermal 
transport may be preferable due to the presence 
of low energy spinon excitations
which could transport heat.  
Another subtle effect is that the superconducting
transition  temperature should be slightly 
smaller if a vison is trapped.  
 
{\bf Practical considerations}:
The experiments proposed above are certainly challenging and would require
a good deal of care.
In tuning out of the superconducting
phase either with temperature or by other means, it is important to make sure
that the sample is definitely no longer superconducting.  This 
could be done by
monitoring the resistance simultaneously.
But better still would be to check directly
that no magnetic flux is still trapped
after the external field is turned off in step (c).

It seems most likely that
the topological order, if present at all in the cuprates,
will be two-dimensional in character.
In that case, it is essential
that the time scale of the 
experiment be smaller than the vison decay time $t_v$. 
Since $t_v$  
increases exponentially with the ratio of the vison gap 
to the temperature, the sample should be heated
just above $T_c$ in step (b).
Moreover, if the vison gap is of order the
pseudogap temperature $T^*$ (eg measured
in photoemission) as we proposed earlier\cite{z2short}, 
it would be preferable to work with 
{\it very} underdoped samples.  This will maximize
the vison decay time $t_v$, being
exponentially large in the 
ratio $T^*$ to $T_c$.

In this paper we have discussed a very general and robust physical effect
that must be present in the underdoped cuprates if they exhibit 
electron fractionalization. The robustness of the effect is due to the
topological order inherent in the fractionalized phase. Experimental
confirmation of this effect would unambiguously establish the 
presence of fractionalization in the underdoped cuprates. Conversely, 
if the experiment fails to observe the
effect when performed with sufficient care, 
it would establish the absence of fractionalization.

Electron fractionalization, and the associated topological order\cite{note2}, 
may well
be more widely prevalent in strongly interacting many-fermion systems than has
been previously assumed. In particular, liquid and solid $He-3$, the two
dimensional electron gas at low density and a plethora of heavy fermion
and organic materials all show a host of unusual phenomena which are poorly
understood.  
Experiments along the lines of those 
proposed in this paper
might enable a 
detection of the otherwise elusive topological
order that may lurk in these systems.    
As in the cuprates, a proximate
superconducting or superfluid phase would be required
to detect - and manipulate - this ``hidden" topological order.

We are grateful to Leon Balents, S. Sachdev, and Doug Scalapino
for fruitful discussions.
This research was generously supported by the NSF 
under Grants DMR-97-04005,
DMR95-28578
and PHY94-07194.

\end{multicols}

\begin{references}
\bibitem{note1} Note that these excitations are distinct from the 
holons and spinons introduced in the slave boson theory of the $t-J$ model 
- in particular they do not carry a $U(1)$ gauge charge which allows them 
to be deconfined in two spatial dimensions. See T. Senthil and 
Matthew P.A. Fisher, condmat/0006500 for a detailed discussion 
of the relationship between the $U(1)$ and $Z_2$ approaches.
\bibitem{note2} Theoretically, a variety of fractionalized phases
with distinct topological orderings is possible (just like in 
the quantum Hall effect).
In this paper, we have focused on the particular kind of fractionalization
that is most relevant to the cuprates. For other materials, other distinct kinds
of topological ordering leading to a different fractionalization pattern are 
possible.
\end{references}
\end{document}